\documentclass[aps,prl,groupedaddress,twocolumn]{revtex4}

\usepackage{graphicx}

\begin{document}

\renewcommand{\vec}[1]{{\mathbf #1}}

\title{Dynamics of Anderson localization in open 3D media}
\author{S.E. Skipetrov}
\email[]{Sergey.Skipetrov@grenoble.cnrs.fr}
\author{B.A. van Tiggelen}
\email[]{Bart.Van-Tiggelen@grenoble.cnrs.fr}
\affiliation{Laboratoire de Physique et Mod\'elisation des Milieux Condens\'es/CNRS,\\
Maison des Magist\`{e}res, Universit\'{e} Joseph Fourier, 38042
Grenoble, France}

\date{\today}

\begin{abstract}
We develop a self-consistent theoretical approach to the  dynamics of Anderson localization in open three-dimensional (3D) disordered media. The
approach allows us to study time-dependent transmission and reflection, and the distribution of decay rates of quasi-modes of 3D disordered
slabs near the Anderson mobility edge.
\end{abstract}


\maketitle

Because of scattering from  heterogeneities, the transmission of a short wave pulse through a disordered medium is strongly delayed. The precise
shape of the decay of the average transmission coefficient $T(t)$ has been extensively studied during the recent years
\cite{watson87,kop97,chabanov03,skip04,cheung04,johnson03}, parallel to related studies of current relaxation in disordered conductors
\cite{alt87,muz95,falko95,smol97,mirlin00} and survival probability decay in classically chaotic open dynamical systems \cite{casati97}. If the
disorder is weak, wave propagation is diffusive and $T(t)$ decays exponentially with time $t$ \cite{watson87,kop97}. This simple result only
holds, however, until the Heisenberg time $t_H = 1/\Delta$, where $\Delta$ is the typical inter-mode spacing. Beyond $t_H$ the decay of $T(t)$
is believed to be slower than exponential due to the so-called ``pre-localized'' states (or modes) that have anomalously long lifetimes
\cite{alt87,muz95,falko95,smol97,mirlin00}. For strong disorder, Anderson localization sets in \cite{anderson58,wiersma97,chabanov00} and the
exponential decay is expected to disappear, at least for quasi-one-dimensional (quasi-1D) samples where $t_H$ becomes
smaller than the arrival time of the pulse. Much less is known about the effect of Anderson localization on $T(t)$ in three-dimensional
(3D) samples, where the Ioffe-Regel criterion of localization $k \ell < 1$ is satisfied or, at least, approached as in recent experiments by
Johnson \textit{et al.} \cite{johnson03} ($k$ is the wavenumber, $\ell$ is the mean free path due to disorder). None of the existing analytic
approaches to Anderson localization (random matrix theory \cite{been97}, nonlinear $\sigma$-model \cite{efetov97,muz95,falko95}, optimal
fluctuation method \cite{smol97}, self-consistent diagrammatic theory \cite{vw80}) have succeeded so far to model the dynamics of Anderson localization in open 3D media.

In the present Letter we apply an improved version of the original self-consistent theory of localization by Vollhardt and W\"{o}lfle \cite{vw80}
to calculate the transmission and reflection coefficients, as well as the distribution of mode decay rates of open 3D disordered media in the
localized regime. The basic idea in this theory is to renormalize the diffusion coefficient to  account for constructive interferences between
reciprocal wave paths inside the sample.
The self-consistent theory of Anderson localization has been recently applied to study
dynamics of weak localization in quasi-1D waveguides and 3D slabs \cite{cheung04},
and dynamics of localized waves in unbounded 1D and 2D media \cite{lobkis05}.
The key distinctive feature of our approach, as opposed to these works, is to allow for
\textit{position} dependence of the renormalized
diffusion coefficient, which is of particular importance near open sample boundaries, where constructive interferences  are strongly suppressed due
to leaks. Our work may guide experimental quests for Anderson localization of light \cite{wiersma97,johnson03} and  be of help in studies of `random lasers' \cite{letokhov67}.

In this paper we consider a 3D disordered slab confined between the planes $z=0$ and $z = L\gg \ell$. Our theoretical model reduces to the
diffusion equation for the intensity Green's function $C$:
\begin{eqnarray}
\left[ -i \Omega - \vec{\nabla} \cdot {\cal D}(\vec{r}, \Omega) \vec{\nabla}
\right] C(\vec{r}, \vec{r}^{\prime}, \Omega) = \delta(\vec{r} - \vec{r}^{\prime})
\label{selfcon1}
\end{eqnarray}
with a position- and frequency-dependent diffusivity ${\cal D}(\vec{r}, \Omega)$
\cite{vw80,tiggelen00}:
\begin{eqnarray}
\frac{1}{{\cal D}(\vec{r}, \Omega)} = \frac{1}{D_B} + \frac{12 \pi}{k^2 \ell}
C(\vec{r}, \vec{r}, \Omega)
\label{selfcon2}
\end{eqnarray}
where $D_B = v \ell/3$ and $v$ is the energy transport velocity.
These equations have to be solved with the boundary conditions
$C \mp z_0 [ {\cal D}(z, \Omega)/D_B ]
\partial_z C = 0$ at $z = 0$ and $z = L$. We adopt $z_0 = 2/3 \ell$. A larger $z_0 \sim \ell$ allows to account
for internal reflections at the surfaces of the slab, but does not affect our conclusions qualitatively.
To regularize the unphysical divergence of $C(\vec{r}, \vec{r}^{\prime}, \Omega)$ for $\vec{r}
\rightarrow \vec{r}^{\prime}$, we work with its Fourier transform ${\hat C}(\vec{q}_{\perp}, z, z^{\prime}, \Omega)$ in the $xy$ plane and then
introduce an upper momentum cut-off $q_{\perp} = \mu/\ell$ to obtain $C(\vec{r}, \vec{r}, \Omega)$ in Eq.\ (\ref{selfcon2}). The numerical
constant $\mu = 1/3$ is chosen such that the localization transition in the infinite medium occurs at $k \ell = 1$.

Following Refs.\ \cite{chabanov03} and \cite{skip04}, we introduce the \textit{leakage function}
\begin{eqnarray}
P_T(\alpha) &=& \frac{i}{2 \pi} \lim_{\epsilon \downarrow 0}
\left[{\cal D}(z,  -i \alpha + \epsilon)
\partial_z {\hat C}(0, z, z^{\prime}, -i \alpha + \epsilon) \right.
\nonumber \\
&-&\left.{\cal D}(z,  -i \alpha - \epsilon)
\partial_z {\hat C}(0, z, z^{\prime}, -i \alpha - \epsilon) \right]
\label{pt}
\end{eqnarray}
with $z = L$ and $z^{\prime} = \ell$.
The Laplace transform of the leakage function yields the time-dependent
transmission coefficient $T(t)$.
Deep in the diffuse regime ($k \ell \rightarrow \infty$) the leakage function
is a sum of delta-peaks located at $\alpha = n^2/t_D$ with
$t_D = (L + 2 z_0)^2/\pi^2 D_B$ the typical time needed to cross the disordered sample by diffusion.
For long times
$t > t_D$ only the first peak is relevant and the transmission coefficient $T(t)$ decays
exponentially with $t/t_D$.

\begin{figure}
\includegraphics[width=8.5cm,angle=0]{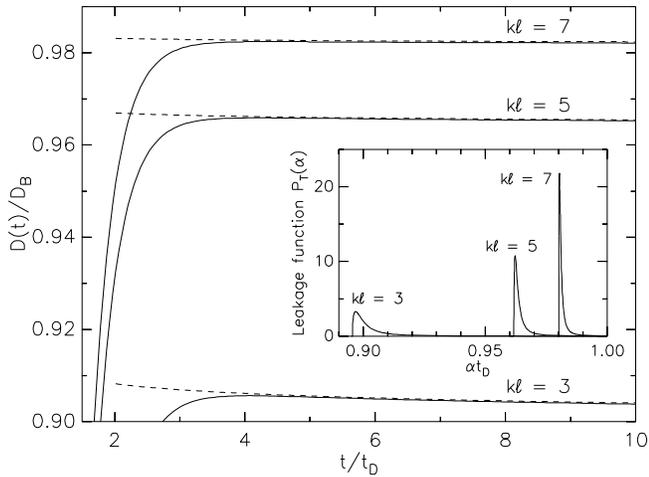}
\caption{\label{fig1} Time-dependent diffusion coefficient for a disordered slab of thickness $L = 100 \ell$ in the regime of
weak disorder (the product of wavenumber $k$ and mean free path $\ell$ exceeds unity). Dashed lines show the result of the perturbational
calculation (\ref{dtweak}). Inset: The first peak of the leakage function $P_T(\alpha)$. We find $P_T(\alpha) \sim \delta(\alpha-1/t_D)$ and
$D(t)/D_B \equiv 1$ in the limit $k \ell \rightarrow \infty$.}
\end{figure}

Let us first see what happens when $k \ell$ is large but finite. We solve Eqs.\ (\ref{selfcon1}) and (\ref{selfcon2}) numerically by extending the approach of Ref.\ \cite{tiggelen00} to $\Omega \ne 0$ and find
that the first peak of $P_{T}(\alpha)$ shifts to smaller values of $\alpha$ while acquiring
a finite width (see the inset of Fig.\ \ref{fig1}). To
characterize the deviation of the transmission coefficient from the pure exponential decay, it has proven convenient \cite{chabanov03,skip04} to
think of $T(t)$ as if it were still resulting from a diffusion process though with a time-dependent diffusion coefficient $D(t) \equiv  -[(L + 2
z_0)^2/\pi^2] (d/dt) \ln T(t)$. As can be seen from Fig.\ \ref{fig1}, $D(t)$ is smaller than $D_B$ and depends on time, though its time
dependence is rather weak. This is also confirmed by the analytic calculation that we performed assuming $t \gg t_D$ and $k \ell \gg 1$:
\begin{eqnarray}
\frac{D(t)}{D_B} &\simeq& 1 - \frac{1}{(k \ell)^2}
\left[ 1 + \frac{3}{2} \frac{\ell}{L}
\left( 7.25 - 4 \ln \frac{L}{\ell} \right) \right]
\nonumber \\
&-& \frac{1}{(k \ell)^2} \frac{3}{2} \frac{\ell}{L}
\left[ \ln \frac{t}{t_D} - \frac{2.25}{t/t_D} \right]
\label{dtweak}
\end{eqnarray}
We see that the time-independent deviation of $D(t)/D_B$ from unity is roughly $1/(k \ell)^2$
(in agreement with Ref.\ \cite{cheung04}, where the dependence of ${\cal D}$ on $\vec{r}$ was ignored),
whereas the time-dependent deviation is another
factor $\ell/L$ smaller. This implies that in 3D samples
non-exponential transmission is much more difficult to observe experimentally than in quasi-1D waveguides,
where a linear decay of $D(t)$ with $t/t_H$ has been predicted and observed in the diffuse regime
$t_D < t_H$ \cite{chabanov03,skip04}. The fact that $D(t)$ is virtually
independent of time in 3D is consistent with purely exponential $T(t)$ observed in
the experiments of Ref.\ \cite{johnson03} despite the rather small value of $k \ell \simeq 4$.

\begin{figure}
\includegraphics[width=8.5cm,angle=0]{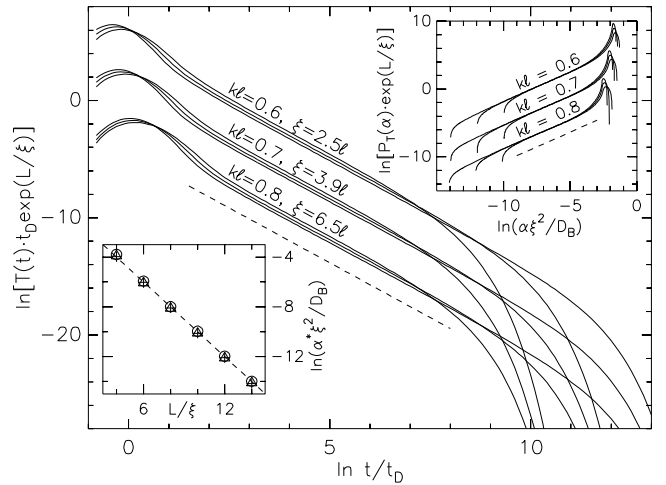}
\caption{\label{fig2} Time-dependent transmission coefficient $T(t)$ of a 3D disordered slab
of thickness $L$ in the localized regime. For each value of $k\ell$,
curves corresponding to $L/\xi= 10$, 12, and 14 are shown ($\xi$ is the localization length).
For clarity, the curves corresponding to $k\ell = 0.7$ and $0.6$ are shifted upwards
by 3 and 6 units, respectively. The $t^{-(1 + s)}$ slope (with $s \approx 0.85$, dashed line) obeyed by all curves crosses over to exponential
decay for times $t > 1/\alpha^*$. Lower inset: leakage threshold $\alpha^*$ versus slab thickness $L$
at $k \ell = 0.6$ ($+$), 0.7 (circles) and 0.8 (triangles). The dashed line is $\alpha^* = (D_B/\xi^2) \exp(-L/\xi)$. Upper inset: leakage
functions $P_T(\alpha)$ behave as $\alpha^s$ (dashed line), except for very small $\alpha < \alpha^*$ where they vanish rigorously.}
\end{figure}

We next turn to the regime of Anderson localization ($k \ell < 1$) which is the primary subject of this paper.
For the parameters used in Fig.\ \ref{fig2} we find the peak of $T(t)$ at
$t \approx t_D$. A detailed study of the arrival time of the pulse will be a subject of our
future work, but we note that
this time is determined by relatively short diffusion paths, corresponding to large-$\Omega$ solutions of Eq.\ (\ref{selfcon1}) and (\ref{selfcon2}), which are only weakly affected by
localization effects. The decay of $T(t)$
following the peak is power-law: $T(t) \sim t^{-(1+s)}$ with $s \simeq 0.85$, in sharp contrast to the exponential decay in the diffuse
regime. The power-law decay of the transmission coefficient originates from the power-law growth of the leakage function $P_T(\alpha) \sim
\alpha^s$ shown in the upper inset of Fig.\ \ref{fig2}. The latter is observed only for $\alpha>\alpha^* = (D_B/\xi^2) \exp(-L/\xi)$ (see the
lower inset of Fig.\ \ref{fig2}), where $\xi = 6 \ell (k \ell)^2/[1 - (k \ell)^4] \ll L$ is the localization length. The leakage function
$P_T(\alpha)$ vanishes for $\alpha<\alpha^*$. An accurate analysis yields $P_T(\alpha) \sim (\alpha - \alpha^*)^p$ (with $p \approx 0.5$) for
$\alpha  \downarrow \alpha^*$, which corresponds to $T(t) \sim \exp(-\alpha^* t)/t^{p + 1}$ for very long times $t > 1/\alpha^*$. The
time-dependent diffusion coefficient
\begin{eqnarray}
\frac{D(t)}{D_B} \simeq \cases{
(s + 1) t_D/t, &$t_D \ll t < 1/\alpha^*$
\cr
\alpha^* t_D + (p + 1) t_D/t, &$t > 1/\alpha^*$
}
\label{dtloc}
\end{eqnarray}
decays as $1/t$, testifying that the transmission coefficient is not pure exponential.
The time window $t_D \ll t < 1/\alpha^{*}$ of power-law decay of $T(t)$
opens up progressively as $\xi$ decreases below $L$. The localization transition at
$k  \ell = 1$ is smooth, in contrast
to the infinite medium where $k \ell = 1$ defines a sharp boundary between diffuse and localized regimes of
wave propagation.

\begin{figure}
\includegraphics[width=8.5cm,angle=0]{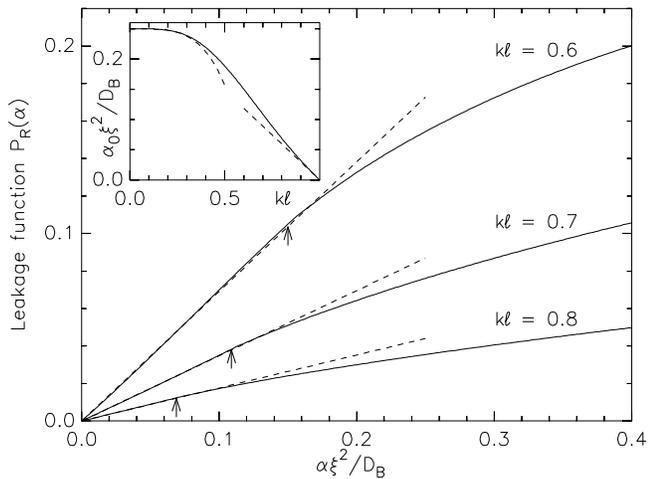}
\caption{\label{fig3} Leakage function $P_R(\alpha)$ in reflection from a 3D disordered half-space in the localized regime. Dashed lines are
linear fits to the initial parts of the curves. Arrows show $\alpha_0$ up to which the linear dependence of $P_R(\alpha)$ on $\alpha$ is
expected. Inset: $\alpha_0$ as a function of $k \ell$. Dashed lines show approximate analytic results $\alpha_0 = (D_B/4 \xi^2)[1 - 6 (k
\ell)^4]$ for $k \ell \downarrow 0$ and $\alpha_0 = (8 D_B/27 \xi^2)(1 - k \ell)$ for $k \ell \uparrow 1$. }
\end{figure}

Let us now consider localization effects in time-dependent reflection, first focusing on a disordered half-space ($L \rightarrow \infty$). For
diffuse waves ($k \ell > 1$), the average reflection coefficient $R(t)$ of the half-space scales as $1/t^{3/2}$. Such a scaling of $R(t)$
corresponds to the leakage function $P_R(\alpha) \sim \sqrt{\alpha}$, where the leakage function
in reflection $P_R(\alpha)$ is defined by Eq.\ (\ref{pt}) with $z = 0$ and with a change of sign.
The square root scaling of $P_R(\alpha)$ is also present in the localized regime, but only for
$\alpha> \alpha_0 \sim D_B/\xi^2$, yielding $R(t) \sim 1/t^{3/2}$ for short times $t \ll 1/\alpha_0$. The waves leaving the sample at such
short times have not penetrated deeply enough to be affected by localization effects and it is therefore natural to recover the diffusion
result. In contrast, $P_R(\alpha$) scales linearly with $\alpha$ for $\alpha < \alpha_0$ (see Fig.\ \ref{fig3}), leading to the scaling law
$R(t) \sim 1/t^2$ for long times $t \gg 1/\alpha_0$. Interestingly, the value of $\alpha_0$ can be calculated analytically, although the
result is quite lengthy (see the inset of Fig.\ \ref{fig3}).

The inverse quadratic scaling of the reflection coefficient with time was reported first for 1D semi-infinite disordered media \cite{white87}.
The transition from $R(t) \sim 1/t^{3/2}$ to $R(t) \sim 1/t^2$ also occurs in quasi-1D disordered waveguides \cite{titov00been,skip04}. The
$1/t^2$ scaling of the reflection coefficient seems therefore to be a hallmark of Anderson localization. For a medium of finite thickness $L \gg
\xi$ this scaling of $R(t)$ should still be valid until a large time $1/\alpha^*$, beyond which an exponential decay sets in.

\begin{figure}
\includegraphics[width=8.5cm,angle=0]{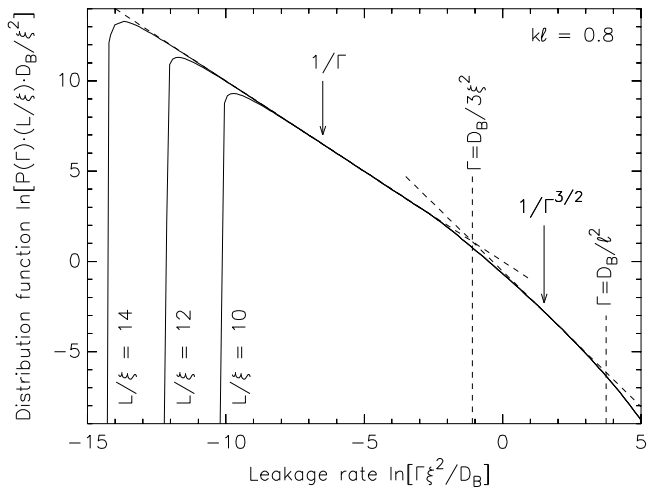}
\caption{\label{fig4}
Distribution of leakage rates $P(\Gamma)$ for a slab of thickness $L$ in the localized regime
(localization length $\xi$). For clarity, only the results for $k \ell = 0.8$ are presented.
For other $k \ell < 1$ the curves are identical, except that the region of
diffuse behavior  $D_B/3 \xi^2 < \Gamma < D_B/\ell^2$ shrinks when $k \ell$ decreases and
$\xi$ approaches $\ell$.}
\end{figure}

An alternative way of dealing with Anderson localization is to study (quasi-)modes of the disordered sample. The statistical distribution
$P(\Gamma)$ of decay rates $\Gamma$ of the modes has recently attracted considerable attention \cite{borgonovi91,casati99,titov00fyod,savin97,kottos},
also in the context of random lasing \cite{patra03}. In the localized regime, the Laplace transform of $P(\Gamma)$
approximately equals the so-called \emph{survival probability} $P_s(t)$ \cite{savin97}.
The latter, in turn, obeys $d P_s(t)/dt = - 2 {\bar T}(t)$, where ${\bar T}(t)$ is the Laplace transform of
the leakage function ${\bar P}_T(\alpha)$ given by Eq.\ (\ref{pt}) averaged over the source position
$z^{\prime}$ inside the disordered sample. 
This yields
$P(\Gamma) = 2 {\bar P}_T(\Gamma)/\Gamma$. The distribution of leakage rates estimated in this way is (see Fig.\ \ref{fig4})
\begin{eqnarray}
&&P(\Gamma) = \\
&&\cases{
0, &$\Gamma < \alpha^*$
\cr
(\xi/L)\times 1/\Gamma,
&$\alpha^* < \Gamma < D_B/3 \xi^2$
\cr
(1/3) \sqrt{D_B/L^2} \times 1/\Gamma^{3/2},
&$D_B/3 \xi^2 < \Gamma < D_B/\ell^2$
\cr
}
\nonumber
\label{pgamma}
\end{eqnarray}

The $1/\Gamma$ behavior of $P(\Gamma)$ is a hallmark of Anderson localization and has already been reported previously for both disordered and
chaotic systems \cite{casati99,kottos} (note that a slightly different result $P(\Gamma) \sim 1/\Gamma^{1.25}$ has been reported  by Titov and
Fyodorov \cite{titov00fyod}). $P(\Gamma) \sim 1/\Gamma^{3/2}$ is typical
for diffuse regime of wave propagation \cite{borgonovi91,kottos}. This
part of $P(\Gamma)$ is due to quasi-modes located closer than $\xi$ to the boundaries of the sample, and which therefore behave more as
extended than as localized modes. $\Gamma > D_B/\ell^2$ correspond to modes with lifetimes
shorter than the mean free time and cannot be correctly described by our theoretical model.

The mode picture of Anderson localization allows a better understanding of $1/t^2$ scaling of short-pulse reflection from a disordered half-space. A short pulse at $z = 0$, $t = 0$ excites
exponentially localized modes of disordered sample with relative weights $\exp(-z/\xi)$, where
$z$ is the position of a given localized mode. Each mode then leaks exponentially with rate
$\Gamma$ proportional to its intensity at the surface of the medium \cite{muz95,smol97}:
$\Gamma \sim \exp(-z/\xi)$. Hence, the relative weight of a given mode is proportional
to its leakage rate $\Gamma$,  which immediately yields $P_{R}(\alpha) \sim \alpha$ and
$R(t) \sim 1/t^2$, if interferences of different modes are ignored.

According to our theory $P_{R,T}(\alpha)$ and $P(\Gamma)$ all exhibit gaps for $0 < \alpha, \Gamma < \alpha^*$. This seems to contradict the
arguments of Refs.\ \cite{muz95,falko95,smol97,kottos} that predict $P(\Gamma) \sim \exp[- \ln^d (\Gamma t_H)]$
(with $d = 2$ or 3) in the limit of small $\Gamma$, due to some special, rare realizations of disorder.
However, we consider a medium with infinite Heisenberg time and
therefore our results cannot be directly compared to those of Refs.\ \cite{muz95,falko95,smol97,kottos}.
The unification of small-$\Gamma$ results corresponding to $t_H \rightarrow \infty$ and
$t_H < \infty$ constitutes a major challenge for future research.

In conclusion, we have presented a self-consistent theoretical approach to the dynamics of
Anderson localization in open 3D disordered media.
We have calculated the time-dependent transmission and reflection, as well as the distribution
of mode decay rates of open disordered slabs in the localized regime.
The transmission exhibits power-law decay in time,
followed by an exponential decay at very large times.
The reflection crosses over from $1/t^{3/2}$ to $1/t^2$ behavior.
The distribution of mode decay rates $P(\Gamma)$ has essentially two parts:
$P(\Gamma) \sim 1/\Gamma$ and $P(\Gamma) \sim 1/\Gamma^{3/2}$,
corresponding to modes localized deep inside the sample and near the boundaries,
respectively.
These results are of particular interest in the context of recent experiments
on Anderson localization of light \cite{chabanov03,johnson03,wiersma97}.

We acknowledge fruitful discussions with Tsampikos Kottos and
Dmitry Savin.



\begin{thebibliography}{99}

\bibitem{watson87}
G.H. Watson, Jr., P.A. Fleury, and S.L. McCall,
\textit{Phys. Rev. Lett.} \textbf{58,} 945 (1987);
J.M. Drake and A.Z. Genack,
\textit{Phys. Rev. Lett.} \textbf{63,} 259 (1989).

\bibitem{kop97}
R.H.J. Kop \textit{et al.,}
\textit{Phys. Rev. Lett.} \textbf{79,} 4369 (1997).

\bibitem{chabanov03}
A.A. Chabanov, Z.Q. Zhang, and A.Z. Genack,
\textit{Phys. Rev. Lett.} \textbf{90,} 203903 (2003).

\bibitem{skip04}
S.E. Skipetrov and B.A. van Tiggelen,
\textit{Phys. Rev. Lett.} \textbf{92,} 113901 (2004).

\bibitem{cheung04}
S.K. Cheung \textit{et al.,}
\textit{Phys. Rev. Lett.} \textbf{92,} 173902 (2004).

\bibitem{johnson03}
P.M. Johnson \textit{et al.,}
\textit{Phys. Rev. E} \textbf{68,} 016604 (2003).

\bibitem{alt87}
B.L. Al'tshuler, V.E. Kravtsov, and I.V. Lerner,
\textit{JETP Lett.} \textbf{45,} 199 (1987);
\textit{Sov. Phys. JETP} \textbf{67,} 795 (1988).

\bibitem{muz95}
B.A. Muzykantskii and D.E. Khmelnitskii, \textit{Phys. Rev. B}
\textbf{51,} 5480 (1995); cond-mat/9601045.

\bibitem{falko95}
V.I. Falko and K.B. Efetov, \textit{Phys. Rev. B} \textbf{52}, 17413 (1995).

\bibitem{smol97}
I.E. Smolyarenko and B.L. Altshuler,
\textit{Phys. Rev. B} \textbf{55,} 10451 (1997).

\bibitem{mirlin00}
A.D. Mirlin, \textit{Phys. Rep.} \textbf{326,} 259 (2000).

\bibitem{casati97}
G. Casati, G. Maspero, and D.L. Shepelyansky,
\textit{Phys. Rev. E} \textbf{56,} R6233 (1997);
K.M. Frahm,
\textit{Phys. Rev. E} \textbf{56,} R6237 (1997);
M. Puhlmann \textit{et al.,}
\textit{Europhys. Lett.} \textbf{69,} 313 (2005).

\bibitem{anderson58}
P.W. Anderson, \textit{Phys. Rev.} \textbf{109}, 1492 (1958);
For recent reviews see Ping Sheng, \emph{Introduction to Wave Scattering,
Localization and Mesoscopic Phenomena} (Academic, San Diego,
1995) or B.A. van Tiggelen, in \emph{Diffuse
Waves in Complex Media}, edited by J.P. Fouque (Kluwer, Dordrecht,
1999), p. 1.

\bibitem{wiersma97}
D.S. Wiersma \textit{et al.,}
\textit{Nature} \textbf{390,} 671 (1997).

\bibitem{chabanov00}
A.A. Chabanov, M. Stoytchev, A.Z. Genack, \textit{Nature}
\textbf{404,} 850 (2000).

\bibitem{been97}
C.W.J. Beenakker, \textit{Rev. Mod. Phys.} \textbf{69,} 1490 (1997).

\bibitem{efetov97}
K.B. Efetov,
\textit{Supersymmetry in Disorder and Chaos} (Cambridge Univ. Press, Cambridge, 1997).

\bibitem{vw80}
D. Vollhardt and P. W\"{o}lfle, \textit{Phys. Rev. B} \textbf{22,}
4666 (1980); in \textit{Electronic
Phase Transitions} (Elsevier Science, Amsterdam, 1992), p. 1.

\bibitem{lobkis05}
O.I. Lobkis and R.L. Weaver,
\textit{Phys. Rev. E} \textbf{71,} 011112 (2005).

\bibitem{tiggelen00}
B.A. van Tiggelen, A. Lagendijk, and D.S. Wiersma, \textit{Phys.
Rev. Lett.} \textbf{84,} 4333 (2000).

\bibitem{letokhov67}
D.S. Wiersma and A. Lagendijk,
\textit{Phys. Rev. E} \textbf{54,} 4256 (1996);
S. Mujumdar \textit{et al.,}
\textit{Phys. Rev. Lett.} \textbf{93,} 053903 (2004).

\bibitem{white87}
B. White \textit{et al.,}
\textit{Phys. Rev. Lett.} \textbf{59,} 1918 (1987).

\bibitem{titov00been}
M. Titov and C.W.J. Beenakker, \textit{Phys. Rev. Lett.}
\textbf{85,} 3388 (2000).

\bibitem{casati99}
G. Casati, G. Maspero, and D.L. Shepelyansky,
\textit{Phys. Rev. Lett.} \textbf{82,} 524 (1999);
M. Terraneo and I. Guarneri,
\textit{Eur. Phys. J. B} \textbf{18,} 303 (2000);
S. Wimberger, A. Krug, and A. Buchleitner, \textit{Phys. Rev. Lett.}
\textbf{89}, 263601 (2002);
F.A. Pinheiro \textit{et al.,}
\textit{Phys. Rev. E} \textbf{69,} 026605 (2004).

\bibitem{titov00fyod}
M. Titov and Y.V. Fyodorov,
\textit{Phys. Rev. B} \textbf{61,} R2444 (2000).

\bibitem{borgonovi91}
F. Borgonovi, I. Guarneri, and D.L. Shepelyansky,
\textit{Phys. Rev. A} \textbf{43,} 4517 (1991);
G. Casati, G. Maspero, and D.L. Shepelyansky,
\textit{Physica D} \textbf{131,} 311 (1999);
A. Ossipov, T. Kottos, and T. Geisel,
\textit{Europhys. Lett.} \textbf{62,} 719 (2003).

\bibitem{savin97}
D.V. Savin and V.V. Sokolov,
\textit{Phys. Rev. E} \textbf{56,} R4911 (1997).

\bibitem{kottos}
T. Kottos, cond-mat/0508173;
M. Weiss, J.A. Mendez-Bermudez, T. Kottos, cond-mat/0509195.

\bibitem{patra03}
M. Patra,
\textit{Phys. Rev. E} \textbf{67,} 016603 (2003).

\end{thebibliography}
\end{document}